\documentclass[11pt]{article}
\usepackage[margin=1in]{geometry}
\usepackage[T1]{fontenc}
\usepackage{mathptmx}
\usepackage{cite}
\usepackage{amsmath}
\usepackage{graphicx}
\usepackage{array}
\usepackage{textcomp}
\usepackage{url}
\usepackage[hidelinks]{hyperref}
\title{Autonomous ETL Pipelines for Salesforce M\&A Consolidation: A Reference Framework, Jitterbit Mapping, and Reproducible Simulation Study}
\author{Imran Basha Shaik\thanks{Corresponding author. Frisco, TX, USA. ORCID: 0009-0002-1210-5646.} \and Sahla Sultana\thanks{Frisco, TX, USA.} \and Khader Moinuddin Shaik\thanks{Frisco, TX, USA.}}
\date{}

\begin{document}
\maketitle

\begin{abstract}
Merging companies that each run Salesforce must consolidate their customer data into a single organization while both source systems remain in use. This requires reconciling different schemas and picklists, resolving duplicate customers across and within organizations, preserving relationships between records, and keeping the target current until cutover. This article proposes a reference framework for an autonomous extract-transform-load pipeline for this task. It combines partitioned bulk extraction, declarative mapping templates, rule-based and statistical quality gates, master data management with deterministic and blocked fuzzy matching, constrained loading with per-record error capture, and change data capture guarded by a data-health circuit breaker. Each stage is mapped to Jitterbit Integration Studio constructs. The framework was implemented as an open reference prototype and evaluated on synthetic two-organization scenarios with known ground truth at three scales, from 32,040 to 383,973 source records, over five random seeds. At a preset threshold, hybrid matching reached pairwise F1 scores of 0.918 for accounts and 0.885 for contacts; calibrating the threshold per object would raise these to 0.938 and 0.963. The quality gates prevented all load failures and all silent loss of unmapped picklist values without wrongly flagging any clean record, whereas removing them produced about 1,000 load failures per run. Over five simulated days of hourly incremental synchronization, replayed events were ignored, an injected upstream fault was contained in the batch in which it occurred, and the target agreed with a full rebuild on more than 99.3\% of field values. Validation on a production deployment remains future work.
\end{abstract}


\section{Introduction}\label{sec:1}

Mergers and acquisitions (M\&A) frequently leave the combined company with two or more customer relationship management (CRM) environments that describe overlapping customers in different ways. When the environments are Salesforce organizations (``orgs''), consolidation into a single target org requires the integration team to reconcile structural differences in objects and fields, translate picklist vocabularies, normalize date, time-zone and contact formats, identify records that describe the same account or person, and re-parent child records so that relationships survive the move. The work must usually be completed while both source orgs remain in daily use, which means the target has to be kept current between the initial bulk load and the final cutover.

Traditional extract-transform-load (ETL) practice was shaped by data warehousing, where sources are comparatively stable and load windows can be scheduled well in advance~\cite{kimball2004,vassiliadis2009}. Consolidation projects stress that model in several ways. Data volumes and the variety of formats have grown~\cite{saggi2018,singh2015}, duplicate detection is a research field in its own right~\cite{christen2012book,elmagarmid2007}, and data quality failures that are caught late are expensive to repair~\cite{batini2009}. Integration Platform as a Service (iPaaS) products such as Jitterbit Harmony provide connectors, mapping tools and schedulers that reduce hand-written code, but they do not by themselves specify how matching, quality control, lineage and incremental synchronization should be combined for a CRM merger.

The academic literature offers mature building blocks for each of these concerns, yet few studies assemble them into an end-to-end design for CRM consolidation and evaluate the assembled design against known answers. Evaluations of consolidation tooling are also hard to reproduce, because merger data are confidential. This paper responds to both problems. Following a design science approach~\cite{hevner2004,peffers2007}, it makes four contributions:

\begin{itemize}
\item A reference framework for autonomous ETL in Salesforce consolidation that combines partitioned extraction, declarative mapping templates, quality gates, master data management (MDM), constrained loading, change data capture (CDC) and data-health monitoring, with explicit control points and audit artefacts (Section~\ref{sec:3}).
\item A mapping of each framework stage onto Jitterbit Integration Studio constructs, so that the design can be implemented on that platform (Section~\ref{sec:3.8}).
\item An open, reproducible simulation that generates two Salesforce-like orgs with differing schemas, known duplicates, deliberately difficult near-duplicates and injected defects, together with ground truth that the pipeline never sees (Section~\ref{sec:4}).
\item An empirical characterization of the framework's behavior, including matching accuracy and its sensitivity to thresholds, the contribution of each quality gate, scaling behavior, and the agreement between incremental synchronization and a full rebuild (Section~\ref{sec:5}).
\end{itemize}

Throughout, we distinguish between what the simulation demonstrates and what would require a production deployment to establish. Section~\ref{sec:6} discusses the findings and Section~\ref{sec:7} sets out the threats to validity.

\section{Related Work}\label{sec:2}

\subsection{ETL architectures and data integration}\label{sec:2.1}

ETL processes extract data from heterogeneous sources, transform them into a common representation and load them into a target. Vassiliadis~\cite{vassiliadis2009} surveys ETL research across conceptual modeling, the individual extraction, transformation and loading stages, and end-to-end concerns such as optimization and resumption. Simitsis et al.~\cite{simitsis2005} treat ETL workflows as state spaces that can be optimized by reordering activities, which supports separating transformation logic from its execution plan. Practitioner texts emphasize staging areas, conformed dimensions and the need for repeatable, auditable jobs~\cite{kimball2004}. Schema matching, the identification of corresponding elements in different schemas, is the precursor to any mapping between two orgs~\cite{rahm2001}.

Big data research extends these ideas to larger and faster-changing sources. Surveys of analytics platforms and distributed frameworks document trade-offs among throughput, latency and fault tolerance and show why monolithic, single-node jobs struggle as volumes grow~\cite{inoubli2018,singh2015}. Cloud architectures add elastic provisioning, which suits the bursty workload of a migration~\cite{khan2013}. Saggi and Jain~\cite{saggi2018} and Alharthi et al.~\cite{alharthi2017} describe how volume, velocity and variety, together with organizational barriers, complicate integration projects. Dong and Srivastava~\cite{dong2015} frame these challenges specifically for data integration, identifying schema alignment, record linkage and data fusion as its core tasks. These are the same three tasks a CRM consolidation must perform.

\subsection{Entity resolution and master data management}\label{sec:2.2}

Entity resolution (also called record linkage or deduplication) identifies records that refer to the same real-world entity. The probabilistic model of Fellegi and Sunter~\cite{fellegi1969} remains the foundation of weighted matching, and Winkler~\cite{winkler1990} introduced the string comparator widely used for names. Elmagarmid et al.~\cite{elmagarmid2007} survey duplicate detection techniques, Christen~\cite{christen2012book} gives a comprehensive treatment of data matching, and Köpcke and Rahm~\cite{kopcke2010} compare matching frameworks. Because comparing every pair of records grows quadratically, practical systems use blocking or indexing to restrict comparisons to plausible candidates~\cite{christen2012survey,papadakis2020}. Christophides et al.~\cite{christophides2020} review end-to-end entity resolution workflows, including incremental settings in which records arrive over time. Gruenheid et al.~\cite{gruenheid2014} show that incremental linkage can diverge from batch linkage unless existing clusters are revisited when new evidence arrives, a point that bears directly on CDC-driven consolidation.

Master data management builds on entity resolution by maintaining a single authoritative (``golden'') record per entity, together with survivorship rules that decide which source supplies each attribute, and a cross-reference from source identifiers to golden identifiers~\cite{loshin2009}. In an M\&A consolidation, that cross-reference is also the lineage record that allows any target value to be traced back to its source.

\subsection{Data quality and automated validation}\label{sec:2.3}

Data quality is multidimensional, spanning accuracy, completeness, consistency and timeliness~\cite{wang1996}. Rahm and Do~\cite{rahm2000} classify data cleaning problems as single-source or multi-source and as schema-level or instance-level, a taxonomy that maps directly onto the defects found when two CRM orgs are merged. Batini et al.~\cite{batini2009} compare methodologies for assessing and improving quality. More recent work treats validation as code: Schelter et al.~\cite{schelter2018} describe declarative ``unit tests for data'' that run at scale, and Breck et al.~\cite{breck2019} describe schema-based validation and drift detection in production pipelines. Roh et al.~\cite{roh2021} review data collection and preparation, stressing that quality controls should sit early in a pipeline rather than after data have been consumed.

\subsection{Change data capture and incremental maintenance}\label{sec:2.4}

Change data capture propagates inserts, updates and deletes from a source instead of repeating full extracts. Kleppmann~\cite{kleppmann2017} discusses log-based CDC, idempotent consumers and the at-least-once delivery semantics that make duplicate events a normal condition rather than an exception. Salesforce exposes change events for supported standard objects and all custom objects through its Change Data Capture feature~\cite{salesforce2026cdc}, and its Bulk application programming interface (API) 2.0 supports asynchronous ingest and query of large datasets~\cite{salesforce2026bulk}. An incremental consolidation pipeline must therefore handle replayed events, maintain checkpoints, and decide what to do with change events that would break integrity rules.

\subsection{Research gap}\label{sec:2.5}

The literature supplies the components of a consolidation pipeline: ETL architecture, schema matching, entity resolution with blocking, survivorship, declarative validation and CDC. What remains under-specified is how these components should be combined, controlled and audited for the particular case of merging Salesforce orgs, and how such a combined design behaves when its outputs can be checked against known answers. Industrial descriptions of iPaaS-based migrations rarely report matching accuracy, the effect of validation gates, or whether incremental synchronization converges to the same state as a full reload. This study addresses that gap with an explicit framework and a reproducible evaluation in which every reported number can be regenerated.

\section{Reference Framework}\label{sec:3}

\subsection{Design objectives and assumptions}\label{sec:3.1}

The framework is designed to satisfy six objectives: (O1) every source record is either represented in the target or accounted for in an audit log; (O2) duplicate entities are merged with measurable precision and recall; (O3) no record that would violate target constraints reaches the load stage; (O4) every target value can be traced to its source records; (O5) the target is kept current between bulk load and cutover, and replayed change events do not alter the result; and (O6) faults in upstream data are detected and contained automatically. The design assumes API access to every source org and to the target, a staging area for intermediate data, and a single acquirer org whose values take precedence when two sources are equally recent.

\begin{figure}[htbp]
\centering
\includegraphics[width=0.98\textwidth]{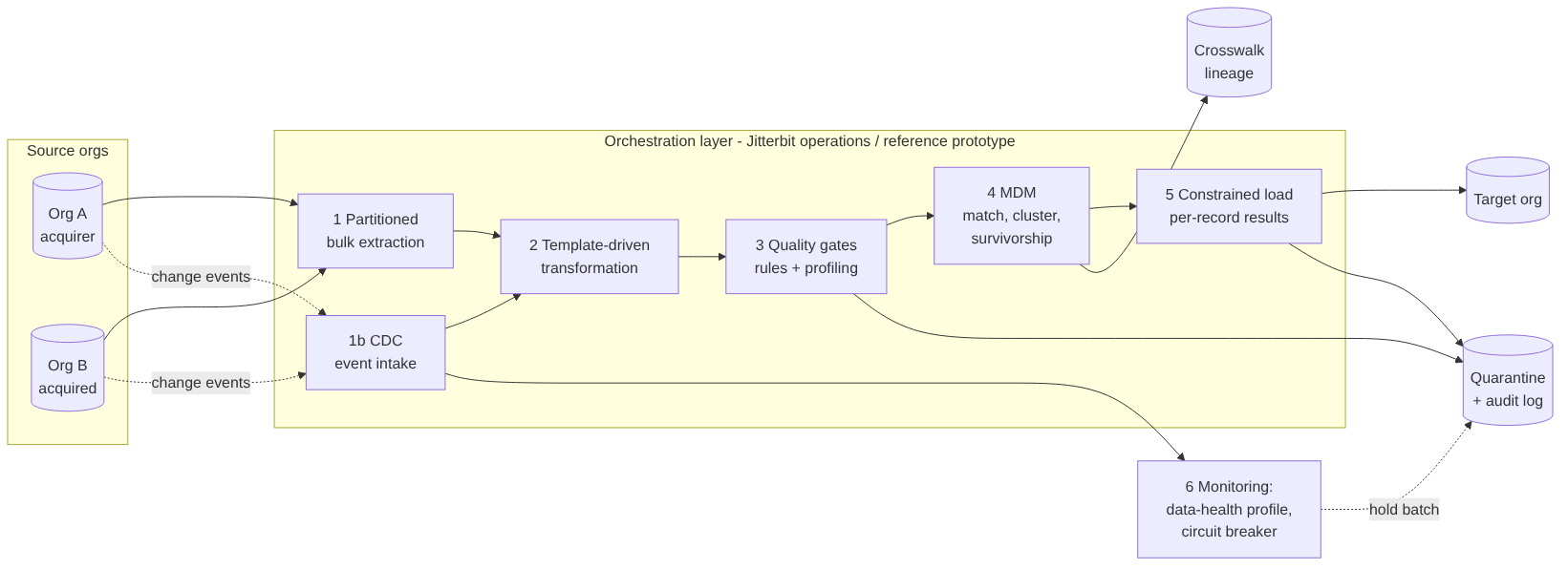}
\caption{Reference architecture. Numbered stages run in the orchestration layer (Jitterbit operations in a production deployment; the Python reference prototype in this study). Rejected and held records go to quarantine with audit entries; the crosswalk records lineage from source to golden identifiers.}
\label{fig:1}
\end{figure}

Fig.~\ref{fig:1} shows the architecture and Table~\ref{tab:1} summarizes the purpose, inputs, outputs and controls of each stage.

\begin{table}[htbp]
\centering
\caption{Pipeline stages and their controls}
\label{tab:1}
\setlength{\tabcolsep}{4pt}
\renewcommand{\arraystretch}{1.15}
\footnotesize
\begin{tabular}{>{\raggedright\arraybackslash\hspace{0pt}}p{0.067\textwidth}>{\raggedright\arraybackslash\hspace{0pt}}p{0.290\textwidth}>{\raggedright\arraybackslash\hspace{0pt}}p{0.223\textwidth}>{\raggedright\arraybackslash\hspace{0pt}}p{0.294\textwidth}}
\hline
\textbf{Stage} & \textbf{Purpose} & \textbf{Main output} & \textbf{Controls} \\
\hline
1 Extraction & Capture a baseline snapshot of each source object & Compressed staged shards partitioned by creation year & Partition keys; parallel workers; shard manifest \\
1b CDC intake & Capture creates, updates and deletes after the baseline & Ordered micro-batches of change events & Event identifiers; idempotent skip of replays \\
2 Transformation & Map each org's schema, picklists and formats to the canonical model & Canonical records with normalized match keys & Versioned templates; unmapped values flagged, never defaulted \\
3 Quality gates & Stop records that would break target rules; flag statistical outliers & Accepted records; quarantine with audit entries & Rule catalog; outlier fence; hold of children whose parent is quarantined \\
4 MDM & Find duplicates and build golden records & Golden accounts and contacts; crosswalk & Deterministic rules; blocking; weighted score threshold; survivorship \\
5 Load & Write golden records and re-parented children & Target records; per-record error log & Batches of 10,000; row-level fallback on failure \\
6 Monitoring & Watch data health of each CDC batch & Alerts; held batches & Profile baseline; z-score threshold; circuit breaker \\
\hline
\end{tabular}
\end{table}

\subsection{Partitioned extraction}\label{sec:3.2}

Each source object is partitioned on its creation date and the partitions are staged in parallel as compressed files with a manifest. Partitioning bounds the size of any single request, allows interrupted extractions to resume at the partition level, and mirrors the way bulk query jobs are typically split to stay within platform limits~\cite{salesforce2026bulk}. The manifest lists each partition, its object and its record count, so that the reconciliation step can confirm that nothing was lost before transformation began.

\subsection{Template-driven transformation}\label{sec:3.3}

Transformation logic is expressed as declarative mapping templates rather than code. A template names, for each source org and object, the source field that supplies each canonical field, the date and time format and time zone of the source, and the versioned picklist map to apply. Org B in the evaluation, for example, stores timestamps as US Central local time in a month-first format, labels the account name field Company\_Name\_\_c, and uses its own industry and stage vocabularies; its template converts all of these to the canonical model. Adding a further org requires a new template, not new code.

Two rules govern transformation. First, a value that cannot be mapped (an unknown picklist value or an unparseable date) is recorded as an issue on the record and passed to the quality gates; it is never silently replaced by a default. Second, the transformation stage computes normalized match keys used later by MDM: company names lower-cased with punctuation and legal suffixes removed, telephone numbers reduced to their last ten digits, web domains stripped of scheme and ``www'', e-mail addresses lower-cased and checked for syntactic validity, and first names mapped through a standard nickname table.

\subsection{Quality gates}\label{sec:3.4}

The quality gates apply a catalog of rules, each recorded in the audit log with the rule name, the field, and the action taken. Blocking rules quarantine a record when a required field is missing, an e-mail address is syntactically invalid, an opportunity amount is negative, a transformation issue was recorded, or a lookup refers to an account that does not exist in any source. A child record whose parent account exists but was itself quarantined is held rather than rejected, so that it can be released once the parent is repaired. Non-blocking rules flag records for review without stopping them. In the prototype, one such rule computes a robust outlier fence on log-transformed annual revenue (upper quartile plus 1.5 times the interquartile range) to catch unit errors that no field-level rule would detect.

\subsection{Master data management}\label{sec:3.5}

MDM runs in two passes, accounts first and then contacts, because contact matching uses the golden account to which each contact belongs.

Matching combines deterministic and probabilistic evidence. Deterministic rules link accounts that share a normalized web domain or the same normalized name and telephone number, and link contacts that share a valid e-mail address. Probabilistic matching then compares candidate pairs produced by blocking. Accounts are blocked on the first four characters of the normalized name, the first five characters of its longest token, and the telephone number. Contacts are blocked within their golden account on the first two characters of the surname and of the first name, and globally on telephone number. Each candidate pair receives a weighted score in the spirit of Fellegi and Sunter~\cite{fellegi1969}: Jaro-Winkler similarity on names~\cite{winkler1990} and exact agreement on city, telephone, domain or e-mail, with weights renormalized over the attributes that both records actually contain. Pairs at or above a threshold $\tau$ are linked, and clusters are formed as connected components of all links.

Survivorship builds one golden record per cluster: for each attribute it takes the most recently modified non-empty value, breaking ties in favour of the acquirer org, and it keeps the earliest creation date. Every source identifier is written to a crosswalk with its golden identifier, which provides the lineage required by objective O4. Opportunities are not deduplicated, because they represent distinct transactions, but their account references are rewritten to golden account identifiers.

\subsection{Constrained load and reconciliation}\label{sec:3.6}

The target schema enforces the constraints that the target org would enforce: required fields, valid lookups, restricted picklists, an e-mail format check and non-negative amounts. Records are loaded in batches of 10,000. If a batch fails, it is retried record by record and each failure is written to an error log, mirroring the per-record results returned by a bulk ingest job. After loading, a reconciliation step confirms that every source record is either represented in the target (directly or through the crosswalk) or present in the audit log, that no lookup in the target points to a missing account, and that the sum of opportunity amounts in the target equals the sum over accepted source opportunities.

\subsection{Change data capture and monitoring}\label{sec:3.7}

After the baseline load, change events are applied in hourly micro-batches. Each event carries an identifier, and any identifier already processed is skipped, which makes the consumer idempotent under at-least-once delivery~\cite{kleppmann2017}. Each event is transformed and validated with the same templates and rules as the bulk path. An update is applied to its source record and the affected golden record is recomputed by survivorship. A new contact is matched against the existing contacts of its golden account (and against the e-mail index) and either joins an existing golden record or creates a new one. Deletions remove the corresponding target record.

Before a batch is applied, the monitoring stage computes a data-health profile, in the prototype the share of contact updates with a blank e-mail address, and compares it with the mean and standard deviation of the first 24 batches. If the batch exceeds the baseline by more than four standard deviations, a circuit breaker holds the entire batch in quarantine for review instead of applying it. This protects the target from an upstream fault, such as a faulty synchronization job that blanks fields, that would pass record-level validation because blank e-mail addresses are individually legal.

\subsection{Mapping to Jitterbit Integration Studio}\label{sec:3.8}

The framework is platform-neutral, but it was designed to be implemented with Jitterbit Harmony. Table~\ref{tab:2} maps each stage to Integration Studio constructs. This mapping is a design proposal: the evaluation in this paper exercises the reference prototype, not a Jitterbit runtime, and specific connector capabilities should be confirmed against the current platform documentation~\cite{jitterbit2026}.

\begin{table}[htbp]
\centering
\caption{Mapping of framework stages to Jitterbit Integration Studio constructs (design mapping, not evaluated)}
\label{tab:2}
\setlength{\tabcolsep}{4pt}
\renewcommand{\arraystretch}{1.15}
\footnotesize
\begin{tabular}{>{\raggedright\arraybackslash\hspace{0pt}}p{0.092\textwidth}>{\raggedright\arraybackslash\hspace{0pt}}p{0.403\textwidth}>{\raggedright\arraybackslash\hspace{0pt}}p{0.403\textwidth}}
\hline
\textbf{Framework stage} & \textbf{Jitterbit construct} & \textbf{Notes} \\
\hline
Extraction & Salesforce connector query or bulk query activities, one operation per object and partition & Partition by CreatedDate or SystemModstamp to bound job size \\
CDC intake & Scheduled operation with incremental query on SystemModstamp, or an external subscriber to Salesforce change events feeding an API endpoint & Persist the last checkpoint and processed event identifiers in project variables or a staging table \\
Transformation & Transformations (mappings) with Jitterbit Script; picklist maps held in lookup tables & Version templates with the project \\
Quality gates & Conditional logic in scripts; failed records routed to a quarantine target & Write rule, field and action to an audit table \\
MDM & Script or external matching service called from an operation; crosswalk written to a staging database & Matching at scale may be better delegated to a database or service \\
Load & Salesforce connector bulk upsert activities using external identifiers & Capture per-record success and failure files \\
Monitoring & Operation schedules, operation actions on failure, and operation logs and alerts in the management console & Profile check runs as the first step of each CDC operation \\
\hline
\end{tabular}
\end{table}

\section{Evaluation Design}\label{sec:4}

\subsection{Research questions}\label{sec:4.1}

The evaluation addresses five research questions (RQs). RQ1: how accurately does the hybrid matcher merge duplicate accounts and contacts, and how much do its deterministic and probabilistic parts each contribute? RQ2: how sensitive is matching to the threshold $\tau$, and what does blocking cost in accuracy? RQ3: which defects do the quality gates catch, which do they miss, and what happens without them? RQ4: how does run time grow with data volume, and which stage dominates? RQ5: does incremental synchronization stay correct under replayed events and upstream faults, and does it converge to the same state as a full rebuild?

\subsection{Synthetic scenario and ground truth}\label{sec:4.2}

No real merger data were available for this study, and real CRM data would in any case carry privacy obligations and lack verified duplicate labels. We therefore built a generator that produces two Salesforce-like orgs with known ground truth. Org A (the acquirer) uses standard field names, ISO-8601 timestamps in Coordinated Universal Time (UTC) and the canonical picklist values. Org B (the acquired company) uses custom field names, US Central local timestamps in a month-first format, and its own vocabulary in which several values map to one canonical value (for example, ``Tech'' and ``Hi-Tech'' both map to Technology), plus a few values with no mapping.

The generator first creates canonical companies and people using the Faker library, then distributes them across orgs: 50\% of companies appear only in Org A, 35\% only in Org B and 15\% in both. A further 4\% of accounts and 3\% of contacts receive a second record within the same org. Copies are perturbed in the ways that make matching difficult in practice: typographical errors; changed or added legal suffixes; ``and'' and ``\&'' interchanged; upper-casing; nicknames and initials; personal instead of work e-mail addresses; missing, reformatted or different telephone numbers; and a different city for 20\% of duplicate accounts. Two kinds of deliberately difficult non-duplicates are added: sibling companies that share a name stem (for example, a ``Holdings'' and a ``Logistics'' company with the same root name), half of which share the parent's web domain, and pairs of people with the same surname at the same account.

Defects are injected at low, realistic rates with their type recorded in the ground truth: invalid e-mail addresses (1.0\% of contacts), missing required names (0.5\% of contacts, 0.3\% of accounts), references to non-existent accounts (0.5\% of contacts and opportunities), negative amounts (0.5\% of opportunities), unmapped picklist values and unparseable dates in Org B (0.5\% and 0.3\%), revenue values inflated a thousand-fold (0.3\% of accounts) and first and last names swapped (0.5\% of contacts). The last two are included because they are plausible-looking and not covered by any field-level rule. The ground truth file is read only by the evaluation code.

\subsection{Experiments and metrics}\label{sec:4.3}

Table~\ref{tab:3} lists the experiments. Matching is scored with pairwise precision, recall and F1 against the true entity labels of the records that entered MDM~\cite{christen2012book}. Quality gates are scored by the share of records carrying each injected defect that were flagged, and by false positives among records with no injected defect. Run time is wall-clock time for the whole batch pipeline. For CDC, events were generated at 2\% of all records per day for five days, spread across hourly batches, with 1\% of events re-delivered to simulate at-least-once delivery. In batch 83 (day 4, 10:00), every event was a contact update in which a simulated faulty upstream job had blanked the e-mail and telephone fields. After the CDC run, the full pipeline was re-run on the final state of the sources (excluding the held batch) and the two targets were compared.

\begin{table}[htbp]
\centering
\caption{Experiments}
\label{tab:3}
\setlength{\tabcolsep}{4pt}
\renewcommand{\arraystretch}{1.15}
\footnotesize
\begin{tabular}{>{\raggedright\arraybackslash\hspace{0pt}}p{0.082\textwidth}>{\raggedright\arraybackslash\hspace{0pt}}p{0.266\textwidth}>{\raggedright\arraybackslash\hspace{0pt}}p{0.358\textwidth}>{\raggedright\arraybackslash\hspace{0pt}}p{0.169\textwidth}}
\hline
\textbf{ID} & \textbf{Question} & \textbf{Data} & \textbf{Repetitions} \\
\hline
E1 & RQ4 scaling & 5,000, 20,000 and 60,000 canonical companies (32,040 to 383,973 source records) & 3 runs per scale; median reported \\
E2 & RQ1 matching and ablation & 10,000 canonical companies (about 64,000 source records) & 5 random seeds \\
E3 & RQ2 threshold sweep and blocking & Same as E2 (sweep); 2,000 companies (blocking) & 5 seeds (sweep); 1 (blocking) \\
E4 & RQ3 quality gates, with and without & Same as E2 & 5 seeds \\
E5 & RQ5 CDC, idempotency, fault containment, equivalence & 20,000 companies (128,037 source records); 120 hourly batches & 1 run \\
\hline
\end{tabular}
\end{table}

The reference prototype and the experiment scripts were written with the assistance of a generative AI coding tool (Claude, Anthropic); the authors reviewed the code and verified every reported number against the raw experiment output. All experiments ran on a Linux container with two CPU cores, Python 3.13, RapidFuzz 3.14 for string similarity and SQLite as the stand-in target. Absolute times therefore describe local computation and exclude network latency and Salesforce API limits.

\section{Results}\label{sec:5}

\subsection{Matching accuracy (RQ1)}\label{sec:5.1}

Table~\ref{tab:4} reports pairwise matching quality at the pre-set threshold $\tau$ = 0.85, averaged over five seeds. On average 11,873 account records and 27,975 contact records entered MDM per seed, containing 2,055 and 2,015 true duplicate pairs respectively.

\begin{table}[htbp]
\centering
\caption{Pairwise matching quality at $\tau$ = 0.85 (mean $\pm$ standard deviation over 5 seeds)}
\label{tab:4}
\setlength{\tabcolsep}{4pt}
\renewcommand{\arraystretch}{1.15}
\footnotesize
\begin{tabular}{>{\raggedright\arraybackslash\hspace{0pt}}p{0.166\textwidth}>{\raggedright\arraybackslash\hspace{0pt}}p{0.187\textwidth}>{\raggedright\arraybackslash\hspace{0pt}}p{0.166\textwidth}>{\raggedright\arraybackslash\hspace{0pt}}p{0.166\textwidth}>{\raggedright\arraybackslash\hspace{0pt}}p{0.166\textwidth}}
\hline
\textbf{Object} & \textbf{Strategy} & \textbf{Precision} & \textbf{Recall} & \textbf{F1} \\
\hline
Account & Deterministic only & 0.902 $\pm$ 0.015 & 0.850 $\pm$ 0.011 & 0.875 $\pm$ 0.011 \\
Account & Fuzzy only & 0.999 $\pm$ 0.001 & 0.769 $\pm$ 0.010 & 0.869 $\pm$ 0.007 \\
Account & Hybrid & 0.906 $\pm$ 0.015 & 0.931 $\pm$ 0.008 & 0.918 $\pm$ 0.011 \\
Contact & Deterministic only & 0.997 $\pm$ 0.003 & 0.694 $\pm$ 0.012 & 0.818 $\pm$ 0.009 \\
Contact & Fuzzy only & 0.999 $\pm$ 0.001 & 0.753 $\pm$ 0.004 & 0.859 $\pm$ 0.003 \\
Contact & Hybrid & 0.997 $\pm$ 0.003 & 0.796 $\pm$ 0.007 & 0.885 $\pm$ 0.004 \\
\hline
\end{tabular}
\end{table}

The hybrid strategy gave the highest F1 for both objects because deterministic and probabilistic evidence find different duplicates. One result was unexpected. For accounts, almost all of the hybrid matcher's precision loss came from the deterministic rule, not from fuzzy matching: fuzzy matching alone was almost perfectly precise (0.999), whereas the shared-domain rule merged subsidiaries that use their parent's web domain. A rule that looks safe because it requires an exact match can therefore be the main source of false merges. For contacts, the remaining errors were almost all missed matches (recall 0.796). In the seed-7 run, every one of the 409 missed contact pairs lacked a shared e-mail address (one record used a personal address or none), and a third also lacked a telephone number on at least one side, leaving the names as the main evidence.

\subsection{Threshold sensitivity and blocking (RQ2)}\label{sec:5.2}

\begin{figure}[htbp]
\centering
\includegraphics[width=0.98\textwidth]{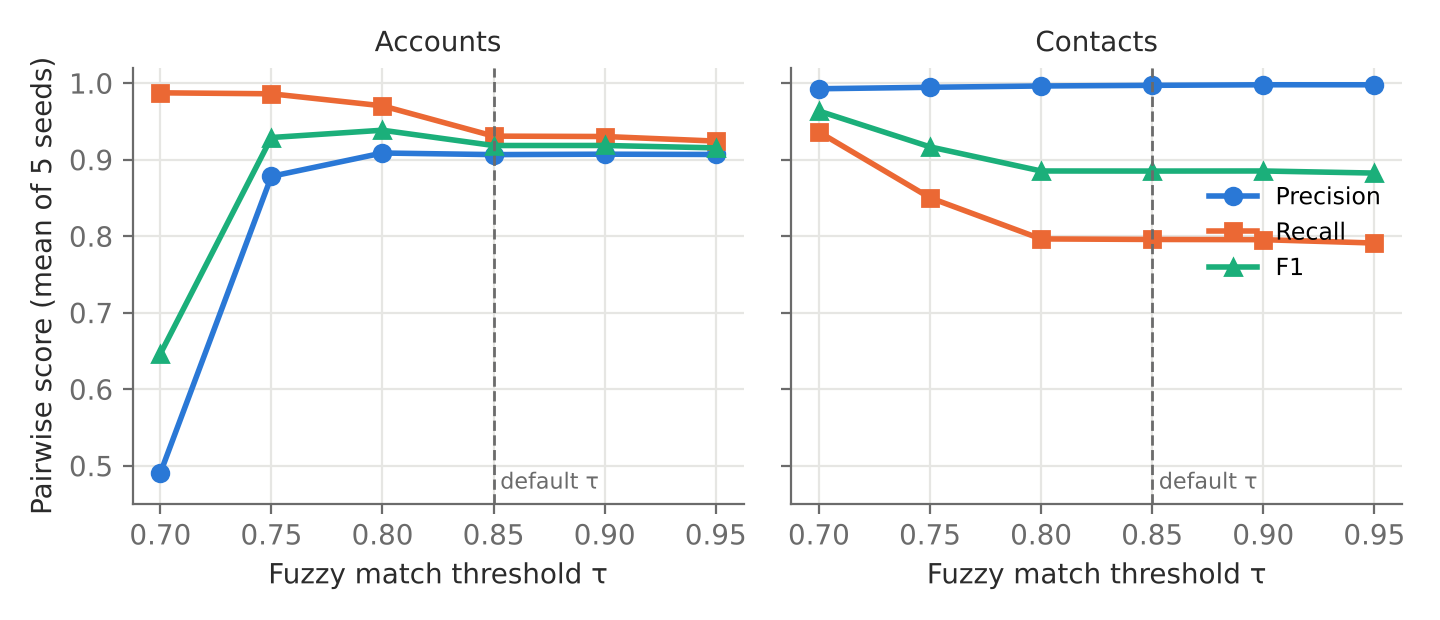}
\caption{Precision, recall and F1 of the hybrid matcher as the fuzzy threshold $\tau$ varies (mean of 5 seeds). The dashed line marks the pre-set value used elsewhere in the evaluation.}
\label{fig:2}
\end{figure}

Fig.~\ref{fig:2} shows that the two objects respond to the threshold very differently. For accounts, F1 peaked at $\tau$ = 0.80 (0.938; precision 0.909, recall 0.970) and collapsed at $\tau$ = 0.70, where precision fell to 0.490. At that threshold, distinct companies that share a leading name token (for example, several firms whose names begin with the same surname) began to link, and because clusters are connected components, a few such links chained many companies together; in the seed-7 run, 886 of 1,038 false pairs shared their first name token. For contacts, precision stayed above 0.99 across the whole range, because blocking within the golden account already excludes most false candidates, so lowering $\tau$ to 0.70 raised recall to 0.936 and F1 to 0.963. A single global threshold is therefore a poor choice: the best thresholds here differ by 0.10 between objects. These best values were selected on the evaluation data and would be optimiztic if reported as expected performance; in practice $\tau$ should be calibrated per object on a labeled sample of the specific migration.

Blocking reduced the number of account pairs compared from 2,767,128 to 22,327 (99.2\% fewer) at the 2,000-company scale, with F1 of 0.910 against 0.912 for exhaustive comparison: one true pair in 392 was lost. At this small scale blocking did not save wall-clock time (0.20 s against 0.19 s), because the exhaustive comparison runs as a single vectorized call, while the blocked version incurs per-block overhead. The benefit of blocking is in avoiding quadratic growth, which matters at the larger scales reported in Section~\ref{sec:5.4}.

\subsection{Quality gates (RQ3)}\label{sec:5.3}

\begin{table}[htbp]
\centering
\caption{Detection of injected defects by the quality gates (5 seeds pooled; 319,606 source records)}
\label{tab:5}
\setlength{\tabcolsep}{4pt}
\renewcommand{\arraystretch}{1.15}
\footnotesize
\begin{tabular}{>{\raggedright\arraybackslash\hspace{0pt}}p{0.441\textwidth}>{\raggedright\arraybackslash\hspace{0pt}}p{0.161\textwidth}>{\raggedright\arraybackslash\hspace{0pt}}p{0.136\textwidth}>{\raggedright\arraybackslash\hspace{0pt}}p{0.136\textwidth}}
\hline
\textbf{Defect type} & \textbf{Records with defect} & \textbf{Flagged} & \textbf{Detection rate} \\
\hline
Invalid e-mail address & 1,390 & 1,390 & 100\% \\
Reference to non-existent account & 1,280 & 1,280 & 100\% \\
Missing required name & 966 & 961 & 99.5\% \\
Negative opportunity amount & 615 & 615 & 100\% \\
Unparseable date & 412 & 412 & 100\% \\
Unmapped picklist value & 366 & 366 & 100\% \\
Revenue inflated $\times$1,000 (outlier rule, non-blocking) & 174 & 174 & 100\% \\
First and last names swapped & 755 & 17 & 2.3\% \\
\hline
\end{tabular}
\end{table}

Table~\ref{tab:5} shows that the gates flagged every record carrying a structural defect that a rule was written for, apart from five records that also had their names swapped: the swap moved the remaining first name into the empty surname field, so the record no longer appeared incomplete. No record without an injected defect was flagged in error (0 of 313,702), and a further 1,698 clean child records were held because their parent account had been quarantined. The statistical fence caught all thousand-fold revenue errors, but swapped names, a semantic error that every field-level rule considers valid, were almost never caught; the 17 that were flagged were caught because of a second defect. Overall, 5,166 of the 5,904 records with an injected defect (87.5\%) were flagged. These detection rates mainly confirm that the rules were implemented correctly, since each rule was written for a defect type that the generator injects; the instructive results are the misses.

Removing the gates showed what they prevent. Without them, an average of 1,032 $\pm$ 45 records per run (about 1.6\% of the source) failed at load, broken down by object as 39 accounts, 657 contacts and 337 opportunities on average, owing to missing required fields, invalid e-mail addresses, missing parents and negative amounts. More seriously, between 39 and 55 opportunities per run were loaded with a blank stage and between 9 and 19 accounts with a blank industry, because unmapped picklist values became empty values that the target accepted. With the gates in place there were no load failures in any run.

\subsection{Scaling (RQ4)}\label{sec:5.4}

\begin{figure}[htbp]
\centering
\includegraphics[width=0.98\columnwidth]{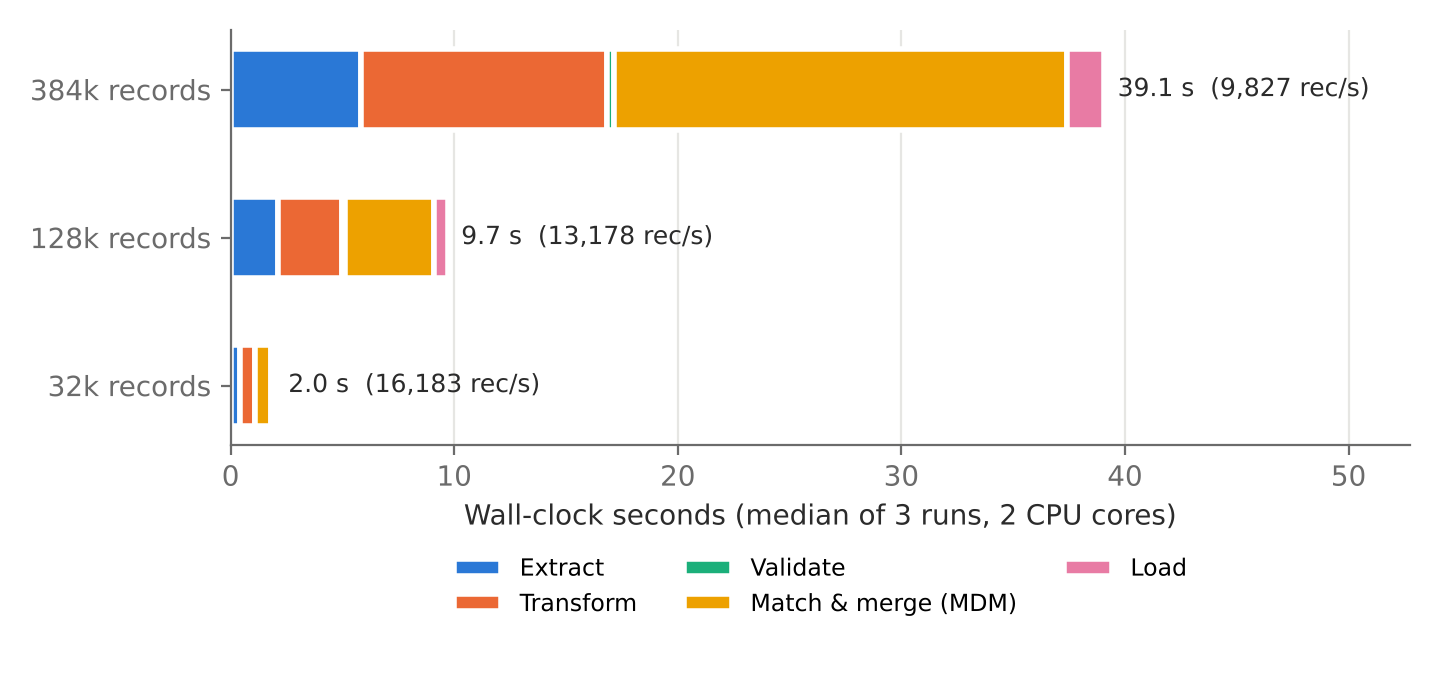}
\caption{Wall-clock time by pipeline stage at three data volumes (median of three runs on two CPU cores; local computation only, excluding network and API latency).}
\label{fig:3}
\end{figure}

\begin{table}[htbp]
\centering
\caption{Batch pipeline results by scale}
\label{tab:6}
\setlength{\tabcolsep}{4pt}
\renewcommand{\arraystretch}{1.15}
\footnotesize
\begin{tabular}{>{\raggedright\arraybackslash\hspace{0pt}}p{0.315\textwidth}>{\raggedright\arraybackslash\hspace{0pt}}p{0.186\textwidth}>{\raggedright\arraybackslash\hspace{0pt}}p{0.186\textwidth}>{\raggedright\arraybackslash\hspace{0pt}}p{0.186\textwidth}}
\hline
\textbf{Measure} & \textbf{32k records} & \textbf{128k records} & \textbf{384k records} \\
\hline
Source records & 32,040 & 128,037 & 383,973 \\
Total time (median of 3) & 2.0 s & 9.7 s & 39.1 s \\
Throughput & 16,183 rec/s & 13,178 rec/s & 9,827 rec/s \\
Account pairs compared & 125,984 & 2,001,702 & 18,910,752 \\
Golden accounts / contacts & 4,963 / 13,319 & 19,847 / 53,157 & 59,571 / 158,894 \\
Records unaccounted for & 0 & 0 & 0 \\
Orphaned lookups in target & 0 & 0 & 0 \\
Opportunity amount checksum & Match & Match & Match \\
\hline
\end{tabular}
\end{table}

Across all three scales, every source record was either represented in the target or recorded in the audit log, no lookup in the target pointed to a missing account, and opportunity amount totals reconciled exactly (Table~\ref{tab:6}). Throughput fell from 16,183 to 9,827 records per second as volume grew twelvefold. Fig.~\ref{fig:3} shows why: MDM grew from 36\% of run time at the smallest scale to 52\% at the largest, because the number of account pairs compared grew roughly 150-fold. The prefix blocks used here become large as the number of companies grows, so blocking keys need tightening, or meta-blocking techniques need applying~\cite{papadakis2020}, before the design is used at production scale. Extraction, transformation and loading grew close to linearly.

\subsection{Change data capture (RQ5)}\label{sec:5.5}

Over 120 hourly batches, 12,840 events were delivered (a mean of 107 per batch). Of these, 84 were re-deliveries and were skipped by event identifier, 12,466 were applied, 184 were rejected by the quality rules (for example, a new contact attached to an account that was in quarantine), and 106 were held by the circuit breaker. Applying a batch took a median of 13.6 ms (95th percentile 18.8 ms; maximum 23.8 ms) of local computation, so in a deployment end-to-end freshness would be governed by the hourly schedule and by API latency rather than by pipeline processing.

\begin{figure}[htbp]
\centering
\includegraphics[width=0.98\columnwidth]{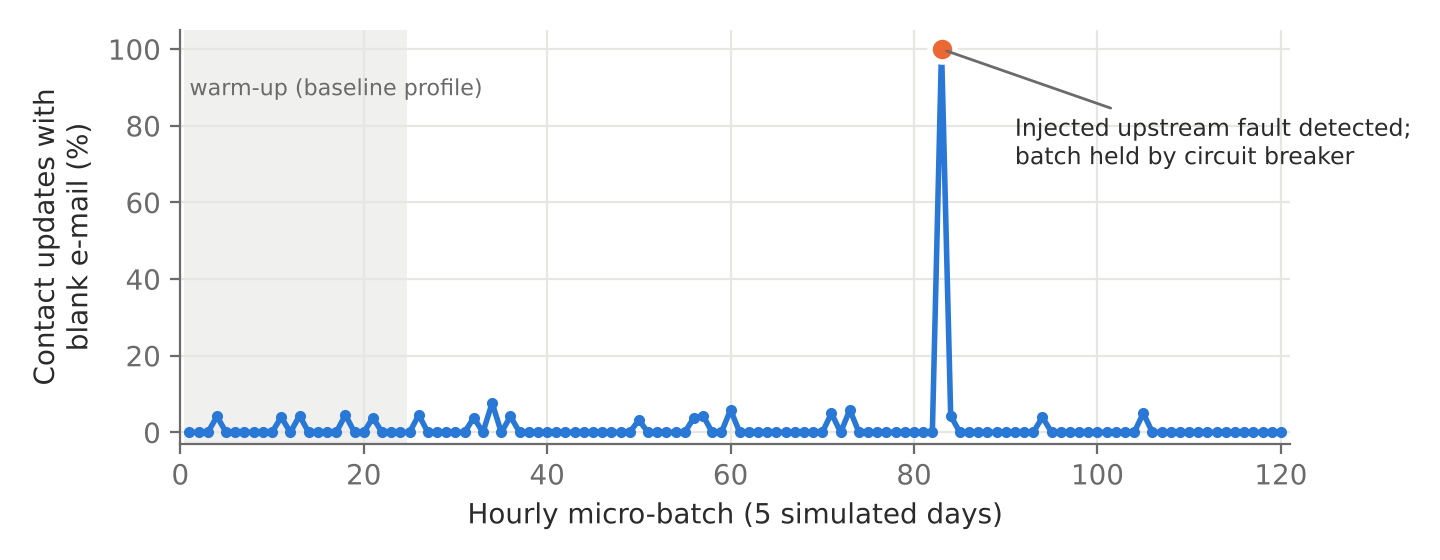}
\caption{Share of contact updates with a blank e-mail address in each CDC batch. The first 24 batches form the baseline profile; the injected upstream fault in batch 83 was detected and held.}
\label{fig:4}
\end{figure}

In normal batches the share of contact updates with a blank e-mail address averaged 0.7\% and never exceeded 7.7\%. In the faulty batch it was 100\%, and the circuit breaker held that batch before any of it was applied (Fig.~\ref{fig:4}). No other batch triggered an alarm, so there were no false alarms over 119 batches.

\begin{table}[htbp]
\centering
\caption{Agreement between the incrementally synchronized target and a full rebuild from the final source state}
\label{tab:7}
\setlength{\tabcolsep}{4pt}
\renewcommand{\arraystretch}{1.15}
\footnotesize
\begin{tabular}{>{\raggedright\arraybackslash\hspace{0pt}}p{0.556\textwidth}>{\raggedright\arraybackslash\hspace{0pt}}p{0.171\textwidth}>{\raggedright\arraybackslash\hspace{0pt}}p{0.171\textwidth}}
\hline
\textbf{Comparison} & \textbf{Records compared} & \textbf{Agreement} \\
\hline
Opportunities: identical set of records & 46,078 & Yes \\
Opportunities: amount and stage values & 46,078 & 100\% \\
Contacts: golden field values per source record & 57,392 & 99.94\% \\
Accounts: golden field values per source record & 23,750 & 99.36\% \\
Contact clusters: pairwise precision against rebuild & 57,392 & 0.994 \\
Contact clusters: pairwise recall against rebuild & 57,392 & 0.999 \\
\hline
\end{tabular}
\end{table}

The incremental target agreed closely, though not perfectly, with a full rebuild (Table~\ref{tab:7}). Opportunities, which involve no matching, agreed exactly. The small differences for contacts and accounts arise because the incremental path matches each new record against existing clusters but never re-clusters existing records when an update changes their matching attributes. An account whose telephone number changes, for example, may join a different cluster in a rebuild. This is the divergence between incremental and batch linkage that Gruenheid et al.~\cite{gruenheid2014} describe. It did not degrade accuracy against the ground truth: after CDC, contact matching F1 was 0.893 for the incremental target and 0.891 for the rebuild. A practical remedy is to run a full rebuild periodically and at cutover, which took 10.4 s at this scale, and to reconcile any differences through the crosswalk.

\section{Discussion}\label{sec:6}

The results support the framework's central design choices while exposing where its defaults should change. Three findings are likely to transfer beyond this simulation.

First, deterministic matching rules need the same scrutiny as probabilistic ones. In this scenario the shared-domain rule caused almost all false account merges: in the seed-7 run, 219 of the 224 false pairs produced by deterministic rules shared a web domain. In real mergers, subsidiaries, franchisees and resellers often share a domain or a main telephone number. Deterministic rules should therefore be treated as strong evidence that feeds a score, not as automatic merges, or they should be restricted to identifiers that are known to be unique, such as tax or registry numbers.

Second, matching thresholds should be calibrated per object. Contact matching benefited from a low threshold because blocking within the golden account had already removed most false candidates, whereas account matching failed abruptly below 0.75. A short labeling exercise on a sample of candidate pairs from the actual orgs, followed by a threshold sweep like that in Fig.~\ref{fig:2}, is inexpensive compared with the cost of false merges, which are harder to undo after cutover than missed matches.

Third, quality gates are most valuable for what the target would not reject. Load-time constraints alone would have rejected about 1,000 records per run, which an operator would see, but they would also have accepted unmapped picklist values as blanks, which an operator would not see. Gates that refuse to default unmapped values, together with reconciliation that accounts for every source record, turn silent data loss into visible, reviewable work. The gates were nevertheless blind to semantic errors such as swapped names. Catching those needs reference data (for example, lists of given names) or learned validators~\cite{breck2019}, and should be part of the review workflow rather than assumed to be covered.

The circuit breaker illustrates a broader point about autonomy. Validating each record individually cannot detect a fault that makes many records individually legal but collectively implausible. Profiling each batch against its own history, and holding rather than applying an anomalous batch, contained the fault automatically without stopping the pipeline. ``Autonomous'' in this framework therefore means that routine decisions (skipping replays, quarantining defective records, holding anomalous batches) are taken without an operator, while every such decision is logged for human review. It does not mean that the pipeline makes judgement calls that cannot be audited.

For practitioners using Jitterbit, the mapping in Table~\ref{tab:2} suggests keeping matching and survivorship in a component that can be tested in isolation, such as a database or matching service called from an operation, rather than spreading the logic across many transformations, and storing the crosswalk and audit log in a queryable staging database that outlives the migration.

\section{Threats to Validity}\label{sec:7}

Construct and external validity are limited chiefly by the use of synthetic data. The generator encodes assumptions about how duplicates and defects arise, and its defect rates, perturbation types and organizational overlap are plausible but not calibrated against a real merger. Real orgs also contain objects and behaviors not modelled here: cases, activities, attachments, record types, sharing rules, and automation (flows and triggers) that may fire during loading. Matching performance on real data could be higher or lower than reported.

The quality-gate detection rates are partly a consequence of design, since the rules were written for defect types that the generator injects. The informative results are the gate ablation and the misses, not the 100\% rates.

Internal validity is supported by separating the ground truth from the pipeline, by fixed random seeds, and by reporting variation across five seeds. The best thresholds in Fig.~\ref{fig:2} were, however, selected on the same data used to report them, and should be treated as an upper bound.

The timing results describe a Python prototype on two CPU cores with SQLite as the target. They do not include network latency, Salesforce API limits or Jitterbit agent performance, and they should not be read as predictions of migration duration on a production platform. The Jitterbit mapping in Table~\ref{tab:2} is a design proposal that has not been executed in this study.

\section{Conclusion and Future Work}\label{sec:8}

This paper proposed a reference framework for autonomous ETL in Salesforce M\&A consolidation, mapped it to Jitterbit Integration Studio, and evaluated an open reference implementation on a synthetic scenario with known ground truth. The hybrid matcher outperformed either of its components; quality gates prevented all load failures and silent loss of unmapped values; reconciliation accounted for every source record at every scale; and CDC synchronization was idempotent under replayed events, contained an injected upstream fault without false alarms, and stayed close to a full rebuild. The evaluation also identified concrete limitations: deterministic rules that cause false merges, the need for per-object threshold calibration, blocking that grows too coarse at scale, blindness to semantic errors, and gradual divergence of incremental clusters.

Future work should (i) implement the framework on a Jitterbit Harmony deployment against Salesforce sandbox orgs to measure API-bound performance; (ii) evaluate it on a real or carefully anonymized merger dataset with manually labeled duplicate pairs; (iii) extend it to activities, cases and record types and to the effect of source automation during loading; (iv) replace fixed rules with learned matchers~\cite{mudgal2018} and incremental re-clustering~\cite{gruenheid2014}; and (v) add reference-data checks for semantic errors.

\section*{Data Availability}
The data generator, pipeline implementation, experiment scripts, and raw results are openly available at \url{https://github.com/imran0547/SDFC/tree/add-autonomous-etl-salesforce-ma-consolidation/autonomous-etl-salesforce-ma-consolidation}. All results reported in this article can be regenerated with the command \texttt{python run\_experiments.py}.

\section*{Competing Interests}
I.~B.~Shaik was previously employed by Salesforce.com.

\section*{Acknowledgment}
Generative artificial intelligence (Claude, Anthropic) was used to assist in writing the simulation code and in drafting and editing the text of this article. The authors reviewed the code, verified every reported result against the experiment output, checked all references, and take full responsibility for the content of the article.

\bibliographystyle{IEEEtran}
\bibliography{refs}
\end{document}